\documentclass[letterpaper,10pt,conference]{ieeeconf}
\IEEEoverridecommandlockouts
\usepackage[sort]{cite}

\usepackage{graphicx}
\usepackage{booktabs}
\usepackage{silence}
\usepackage{subcaption}

\usepackage[T1]{fontenc}

\usepackage{mathtools,amsfonts,amssymb,dsfont}
\usepackage[svgnames]{xcolor}

\newcommand*\bolditalic[1]{\textbf{\textit{#1}}}

\usepackage[standard,amsmath]{ntheorem}

\usepackage{tikz,tikz-cd}
\usetikzlibrary{positioning}

\tikzset{state/.style={circle, draw,
                line width=1bp,
                inner sep=0.25em,
                draw=black!85!white,
                minimum width=2em,
                fill=gray!25!white,
                }}

\newcommand*\ALPHABET{\mathcal}
\newcommand\reals{\mathds{R}}

\newcommand\naturalnumbers{\mathds{N}}
\newcommand\PR{\mathds{P}}
\newcommand\EXP{\mathds{E}}
\newcommand\IND{\mathds{1}}
\newcommand\GRAD{\nabla}

\newcommand*\sublabel[1]{\textup{\textsc{\MakeLowercase{#1}}}}

\newcommand\PiHND{\vec \Pi_{\sublabel{ND}}}
\newcommand\PiHNS{\vec \Pi_{\sublabel{NS}}}
\newcommand\JHND{\vec J^\star_{\sublabel{ND}}}
\newcommand\JHNS{\vec J^\star_{\sublabel{NS}}}

\newcommand\DRD{\mathcal{D}_{\sublabel{D}}}
\newcommand\DRS{\mathcal{D}_{\sublabel{S}}}
\newcommand\XIS{\Delta(\ALPHABET S \times \ALPHABET Z)}

\newcommand\PiZND{\Pi_{\sublabel{ND}}}
\newcommand\PiZNS{\Pi_{\sublabel{NS}}}
\newcommand\PiZSD{\Pi_{\sublabel{SD}}}
\newcommand\PiZSS{\Pi_{\sublabel{SS}}}
\newcommand\JZND{J^\star_{\sublabel{ND}}}
\newcommand\JZNS{J^\star_{\sublabel{NS}}}
\newcommand\JZSD{J^\star_{\sublabel{SD}}}
\newcommand\JZSS{J^\star_{\sublabel{SS}}}
\newcommand\piZND{\boldsymbol \pi^\star_{\sublabel{ND}}}
\newcommand\piZNS{\boldsymbol \pi^\star_{\sublabel{NS}}}
\newcommand\piZSD{\boldsymbol \pi^\star_{\sublabel{SD}}}
\newcommand\piZSS{\boldsymbol \pi^\star_{\sublabel{SS}}}

\newcommand*\IS{\sublabel{IS}}
\newcommand*\Des{\sublabel{Des}}
\newcommand*\PROD{\sublabel{Prod}}
\newcommand*\AIS{\sublabel{AIS}}
\newcommand*\ASQL{\sublabel{ASQL}}
\newcommand*\ASAC{\sublabel{ASAC}}

\newcommand*\F{\mathfrak{F}}
\newcommand\FTV{\F_{\mathrm{TV}}}
\newcommand\dTV{d_{\mathrm{TV}}}

\newcommand\FWas{\F_{\mathrm{Was}}}
\newcommand\dWas{d_{\mathrm{Was}}}

\DeclareMathOperator{\Lip}{Lip}
\DeclareMathOperator{\Span}{span}

\usepackage{tcolorbox}
\tcbuselibrary{theorems,breakable,skins}
\newtcbtheorem[auto counter]{colortheorem}%
  {Theorem}
  {fonttitle=\bfseries\upshape\vphantom{$\Big($}, fontupper=\slshape,
   enhanced,
   drop lifted shadow,
   colbacktitle=black, 
   coltitle=gray!20,
   breakable,
   boxrule=1pt,
   colframe=black, %
   left=2mm,
   right=2mm,
   boxsep=0pt}{thm}

\tcbset{colorformula/.style={
   enhanced,
   drop lifted shadow,
   coltitle=gray!20,
   boxrule=1pt,
   colframe=black, %
   }}

\tcbset{every box/.style={highlight math style={
   enhanced,
   drop small lifted shadow,
   coltitle=gray!20,
   left=1mm,
   right=1mm,
   top=1mm,
   bottom=1mm,
   boxrule=1pt,
   colframe=black, %
   }}}

\makeatletter
\let\NAT@parse\undefined
\makeatother
\usepackage{hyperref}
\hypersetup{
    colorlinks=true,
    linkcolor=blue,
    filecolor=magenta,      
    urlcolor=cyan,
    citecolor=Green,
    }

\makeatletter
\newcommand\tcb@cnt@colortheoremautorefname{Theorem}
\makeatother

\newcommand{\appref}[1]{\hyperref[#1]{Appendix~\ref*{#1}}}

\title{Agent-state based policies in POMDPs: Beyond belief-state MDPs}
\author{Amit Sinha and Aditya Mahajan%
\thanks{The authors are with the Department of Electrical and Computer Engineering, McGill University, Montreal, QC, Canada. They are also affiliated with CIM, GERAD, Mila, and ILLS. Emails: amit.sinha@mail.mcgill.ca, aditya.mahajan@mcgill.ca}%
\thanks{This research was supported in part by NSERC Alliance International Catalyst Grant ALLRP 580801-2022 and in part by a grant from Google's Institutional Research Program in collaboration with Mila.}}

\begin{document}
\raggedbottom

\maketitle

\begin{abstract}
  The traditional approach to POMDPs is to convert them into fully observed MDPs by considering a belief state as an information state. However, a belief-state based approach requires perfect knowledge of the system dynamics and is therefore not applicable in the learning setting where the system model is unknown. Various approaches to circumvent this limitation have been proposed in the literature. We present a unified treatment of some of these approaches by viewing them as models where the agent maintains a local recursively updateable ``agent state'' and chooses actions based on the agent state. We highlight the different classes of agent-state based policies and the various approaches that have been proposed in the literature to find good policies within each class. These include the designer's approach to find optimal non-stationary agent-state based policies, policy search approaches to find a locally optimal stationary agent-state based policies, and the approximate information state to find approximately optimal stationary agent-state based policies. We then present how ideas from the approximate information state approach have been used to improve Q-learning and actor-critic algorithms for learning in POMDPs.
\end{abstract}

\section{Introduction}

Partially observable Markov decision processes (POMDPs) are a widely used model for the optimal control of dynamical systems with partial state observation. They have been extensively studied across various research communities including systems and control, operations research, and artificial intelligence. 

A key conceptual challenge for POMDPs is that the data available at the agent---the history of observations and actions---is increasing with time. The standard approach is to compress this increasing data into a finite dimensional statistic known as the belief state, which is the posterior density of the unobserved state conditioned on the history of observations and actions and, therefore, may be viewed as a generalization of non-linear filtering of controlled processes. 
The belief state is sufficient for evaluating the per-step reward, can be updated recursively, and is strategy independent. Therefore, we can write a dynamic programming decomposition using the belief state as an information state~\cite{Astrom1965,Smallwood1973,Cassandra1994, krishnamurthy2016partially}. Furthermore, the value function of the corresponding belief-state MDP has certain qualitative properties (it is piecewise linear and convex), which can be leveraged for efficient computational algorithms. The earliest such algorithm was the ``one-pass'' algorithm by Smallwood and Sondik~\cite{Smallwood1973}. Various efficient refinements of this algorithm have been presented in the literature, including linear-support algorithm~\cite{Cheng1988}, witness algorithm~\cite{Cassandra1994}, incremental pruning~\cite{Zhang1996, Cassandra1997}, point-based methods~\cite{Pineau2003,Smith2004,Spaan2005}, and others. See~\cite{hauskrecht2000value,kochenderfer2022algorithms} for a unified overview of the numerical methods.

There are two limitations of the belief-state based approach. First, implementing a belief-state based policy is computationally challenging. Keeping track of the belief state requires non-linear filtering, which can be approximated via particle filtering, but is still computationally heavy and is therefore difficult to implement in embedded hardware in robotics and other applications. Second, the belief state is model dependent. Therefore, it cannot be used in model-free reinforcement learning (RL).

An alternative approach, which is more amenable to the learning setting, is to consider what we call the \bolditalic{agent state}. In particular, we relax the assumption that the agent can use the entire history of observations and actions (or a model-dependent compression of it like the belief state) to make a decision. Rather, we assume that the agent maintains a local state that does not depend on the model and makes a decision as a function of the agent state. Such an agent-state formulation has been proposed multiple times in the literature~\cite{sondik1978optimal,Meuleau1999,hutter2005universal,subramanian2022approximate,Dong2022,lu2023reinforcement}. Perhaps the simplest example of the agent state is an agent keeping track of a finite window of past observations and actions, which was first considered in~\cite{Platzman1977, white1994finite} and is commonly referred to as \emph{frame stacking} in the RL literature~\cite{mnih2013playing}. It is argued in~\cite{Dong2022,lu2023reinforcement} that in many of the popular implementations of RL algorithms, the agent state can be further decomposed into three parts: an algorithmic state which is data to be used by subsequent computations, a situational state, which is a summary of the agent's current situation, and an epistemic state, which is the summary of the agent's current knowledge of the environment. We do not pursue such a distinction here but, roughly speaking, our notion of agent state is similar to the situational state in~\cite{Dong2022,lu2023reinforcement}.

\subsection*{Notation}
We use uppercase letters to denote random variables (e.g. $S,A$, etc.), lowercase letters to denote their realizations (e.g. $s,a$, etc.) and calligraphic letters to denote sets (e.g. $\ALPHABET{S}, \ALPHABET{A}$;ggh, etc.). Subscripts (e.g. $S_t,A_t$, etc.) denote variables at time~$t$. $\Delta (\ALPHABET{S})$ denotes the space of probability measures on a set $\ALPHABET{S}$; $\PR(\cdot)$ and $\EXP[\cdot]$ denote the probability of an event and the expectation of a random variable, respectively; and $\IND$ denotes the indicator function.

\subsection*{Organization}
The rest of the paper is organized as follows. In \autoref{sec:model} we present the mathematical model of POMDPs, introduce the agent-state framework, and present sufficient conditions for the agent state to be an information state. 
In \autoref{sec:agent-state} we discuss the different classes of agent-state policies and present the three approaches for finding optimal policies among different policy classes. In \autoref{sec:RL} we discuss various RL approaches taken for learning agent-state based policies. Finally, we present concluding remarks and discussions in \autoref{sec:conclusion}.

\section{The POMDP model and agent-state based policies}\label{sec:model}
\subsection{System model}
Consider a stochastic dynamical system with state $S_t \in \ALPHABET S$, input $A_t \in \ALPHABET A$, and output $Y_t \in \ALPHABET Y$. To simplify the discussion, we will assume that all sets are finite valued and ignore integrability and measurability issues, existence of suprema, etc. See the companion paper~\cite{serdar} for a nuanced discussion of these issues.  The system operates in discrete time with the dynamics given as follows: 
for any time $t \in \naturalnumbers$, we have
\begin{align*}
    \PR(S_{t+1}, Y_{t+1} \mid S_{1:t}, Y_{1:t}, A_{1:t})
    &=
    \PR(S_{t+1}, Y_{t+1} \mid S_{t}, A_{t})
    \\
    &\eqqcolon P(S_{t+1}, Y_{t+1} \mid S_t, A_t)
\end{align*}
where $P$ is a probability transition matrix.\footnote{For the simplicity of notation, we are using a slightly informal notation. Terms such as $\PR(S_{t+1}, Y_{t+1} \mid S_t, A_t)$ should either be viewed as the numerical value $\PR(S_{t+1} = s_{t+1}, Y_{t+1} = y_{t+1} \mid S_t = s_t, A_t = a_t)$ for specific realizations $(s_{t+1},y_{t+1},s_t,a_t)$ of $(S_{t+1},Y_{t+1},S_t,A_t)$ or as probability mass function $\PR(S_{t+1} = \cdot, Y_{t+1} = \cdot \mid S_t = s_t, A_t = a_t)$ or as a distribution-valued random variable $\PR(S_{t+1} = \cdot, Y_{t+1} = \cdot \mid S_t, A_t)$. Typically, all of these interpretations are consistent. If a specific interpretation is needed, we will use a more elaborate notation as appropriate.} In addition, at each time the system yields a reward $R_t = r(S_t, A_t)$. We will assume that $R_t \in [0, R_{\max}]$. 

There is an agent (also called a controller or a decision maker, depending on the research community) which observes the outputs of the system and chooses control actions as inputs to the system. In principle, this agent can be as sophisticated as we want and can, therefore, use the entire history of past observations and actions to choose its action, i.e.,\footnote{Again, we are using a slightly informal notation. We can interpret the above either as $a_t = \vec \pi_t(y_{1:t}, a_{1:t-1})$ for special realizations $(y_{1:t}, a_{1:t})$ of $(Y_{1:t},A_{1:t})$ or as an equality between random variables $A_t(\omega)$ and $\vec \pi_t(Y_{1:t}(\omega), A_{1:t-1}(\omega))$.}

\[
    A_t = \vec \pi_t(Y_{1:t},A_{1:t-1}),
    \quad t \in \naturalnumbers
\]
where $\vec \pi_t$ is called the control law or decision rule at time~$t$. We are using the vector accent to highlight the fact that the policy is history dependent.

We assume that the system runs for an infinite horizon and use $\boldsymbol{\vec \pi} = (\vec \pi_1, \vec \pi_2, \dots)$ to denote the control policy (or simply the policy).\footnote{We are making a deliberate choice of using bold $\boldsymbol{\vec \pi}$ to denote the policy to distinguish between control laws and control policies.} Let $\PiHND$ denote the set of all history dependent (indicated by the vector accent) non-stationary (i.e., time-varying) and deterministic policies.

We assume that the initial state $S_1$ is distributed according to probability mass function $\xi_1$. Then, the performance of any policy $\boldsymbol{\vec \pi} \in \PiHND$ is given by
\[
    J^{\boldsymbol{\vec \pi}} \coloneqq
    \EXP^{\boldsymbol{\vec \pi}}\biggl[ \sum_{t=1}^{\infty} \gamma^{t-1} R_t \biggm| S_1 \sim \xi_1 \biggr]
\]
where $\gamma \in (0,1)$ is the discount factor. Let $\JHND$ (again, the vector accent highlights that we are optimizing over all history dependent policies) denote the optimal performance in $\PiHND$, i.e.,
\[
    \JHND \coloneqq \sup_{\boldsymbol{\vec \pi} \in \PiHND} J^{\boldsymbol{\vec \pi}}.
\]

\subsection{Some remarks on the model}
\begin{enumerate}
    \item The system described above is referred to as a \emph{partially observed} Markov decision process (POMDP) to highlight the fact that the agent sees \emph{partial} observations of the state of the environment. 
    \item Since the per-step reward is uniformly bounded, $J^{\boldsymbol{\vec \pi}}$ is well defined. However, it is not immediately obvious that there exists an optimal policy $\boldsymbol{\vec \pi}^\star$ such that $\JHND =     J^{\boldsymbol{\vec \pi}^\star}$. 
    \item There are several technical questions that need to be resolved carefully when the variables are continuous valued. We refer the reader to the companion paper~\cite{serdar} for a detailed discussion.
    \item In the literature, it is often assumed that
    \begin{align*}
        &\PR(S_{t+1}, Y_{t+1} \mid S_t, A_t) 
        \\ & \quad 
        = \PR(S_{t+1} \mid S_t, A_t) \PR(Y_{t+1} \mid S_{t+1}, A_t)
    \end{align*}
    or sometimes even
    \begin{align*}
        &\PR(S_{t+1}, Y_{t+1} \mid S_t, A_t) 
        \\ &\quad
        = \PR(S_{t+1} \mid S_t, A_t) \PR(Y_{t+1} \mid S_{t+1}).
    \end{align*}
    Such an assumption is not needed for the discussion presented in this paper.

    \item In the discussion above, we have restricted attention to deterministic policies. In principle, we could have also considered non-stationary history dependent stochastic policies $\boldsymbol{\vec \pi} = (\vec \pi_1, \vec \pi_2, \dots)$, where $\vec \pi_t \colon \ALPHABET H_t \to \Delta(\ALPHABET A)$.  Let $\PiHNS$ denote the set of all non-stationary history dependent stochastic policies. Define
    \[
        \JHNS \coloneqq \sup_{\boldsymbol{\vec \pi} \in \PiHNS} J^{\boldsymbol{\vec \pi}}.
    \]
    By definition $\PiHND \subseteq \PiHNS$; hence, $\JHND \le \JHNS$.  However, since the agent has perfect recall (i.e., remembers everything that it has seen and done in the past)
    it can be shown that \bolditalic{there is no loss of optimality in restricting attention to deterministic strategies,}%
    \footnote{\label{fnt:Kuhn}To explain the high-level idea of the result, we borrow the terminology of pure, mixed,  and behavioral strategies used in game theory. A pure strategy is what we call deterministic policy; a behavioral strategy is what we call stochastic policy. A mixed strategy is a probability distribution over pure strategies where the agent picks a pure strategy at the beginning of the game according to specified probability distribution and then follows the pure strategy throughout the game. With this terminology, the result follows from two facts. First, since we are in a single agent unconstrained expectation maximization setting, mixed and pure strategies have the same performance. Second, since we are in a perfect recall setting,  mixed and behavioral strategies have the same performance due to Kuhn's theorem (see e.g., \cite{aumann1961mixed}).}
    i.e., 
    \begin{equation}\label{eq:history}
        \tcbhighmath{\JHND = \JHNS.}
    \end{equation}
\end{enumerate}

\subsection{Agent-state based policies}

We now describe the agent-state based approach, where it is assumed that instead of using the entire history of observations and actions to make a decision the agent maintains a local state (which we will refer to as the agent state and denote by $Z_t \in \ALPHABET Z$) and makes a decision as a function of the agent state. The agent starts with an initial state $Z_1 = \phi_0(Y_1)$ and recursively updates it as follows:\footnote{In principle, the update can be stochastic and be given by 
\( Z_{t+1} = \phi(Z_t, Y_{t+1}, A_t, W_t) \)
where $\{W_t\}_{t \ge 1}$ is a sequence of i.i.d.\ (independent and identically distributed) randomizing variables that are independent of all other primitive random variables.}
\begin{equation}\label{eq:Z-update}
    Z_{t+1} = \phi(Z_t, Y_{t+1}, A_t), \quad \forall t \in \naturalnumbers.
\end{equation}
We call $\phi_0$ the \emph{state-initialization function} and $\phi$ the \emph{state-update function}. 
The agent chooses an action either using a deterministic control law $\pi_t \colon \ALPHABET Z \to \ALPHABET A$ as
\[
 A_t = \pi_t(Z_t)
\]
or using a stochastic control law $\pi_t \colon \ALPHABET Z \to \Delta(\ALPHABET A)$ as
\[
 A_t \sim \pi_t(Z_t).
\]
We call such a policy as an \bolditalic{agent-state based policy}.

A simple example of agent-state based policy is when the agent uses a finite window of past observations and actions, i.e., $Z_t = (Y_{t-n:t}, A_{t-n:t-1})$. Such a model is sometimes called frame-stacking in the RL literature \cite{mnih2013playing, white1994finite}. Another example is when the controller is a finite state automaton~\cite{Sandell1974,Platzman1977,Hansen1998}.
It is not necessary for the agent state to be finite. In fact, the belief-state representation is a special case of agent-state model where the agent state belongs to a $|\ALPHABET S|$-dimensional simplex. However, for the convenience of notation, we would assume that $\ALPHABET Z$ is finite. The discussion can be generalized to continuous valued $\ALPHABET Z$ under appropriate technical assumptions.

There are three fundamental questions when working with an agent state:
\begin{enumerate}
    \item[(Q1)] When is there no loss of optimality in restricting attention to agent-state based policies?
    \item[(Q2)] For a given agent-state update function $\phi$, how do we find the optimal agent-state based policy?
    \item[(Q3)] For a given agent-state space $\ALPHABET Z$, what is the optimal agent-state update function and agent-state based policy?
\end{enumerate}

The short answer to (Q1) is simple: \bolditalic{if an agent state is an information state, then there is no loss of optimality in restricting attention to an agent-state policy.} Of course, this answer only makes sense if we define an information state, which we do in the next section. The answers to (Q2) and (Q3) are more difficult and depend on what we mean by ``optimal''. We will discuss the different conceptual approaches that have been used in the literature in Sec.~\ref{sec:agent-state}.

\subsection{Information state}\label{sec:info-state}

We start with some notation. Let $H_t = (Y_{1:t}, A_{1:t-1})$ denote the history of observations and actions of the agent up to time~$t$ and let $\ALPHABET H_t$ denote the space of realization of all such histories. We can recursively unroll the agent-state update function~\eqref{eq:Z-update} and define a sequence of functions $\boldsymbol{\vec \sigma} \coloneqq (\vec \sigma_1, \vec \sigma_2, \dots)$, where  $\vec \sigma_t \colon \ALPHABET H_t \to \ALPHABET Z$, such that $Z_t = \vec \sigma_t(H_t)$. In particular, 
\begin{equation}\label{eq:history-compression}
    \vec \sigma_1(H_1) = \phi_0(Y_1),
    \hskip 0.5em
    \vec \sigma_2(H_2) = \phi(\vec \sigma_1(H_1),Y_2,A_1), 
    \hskip 0.5em \text{etc.}
\end{equation}
We call $\boldsymbol{\vec \sigma}$ to be the \emph{history compression function} corresponding to the agent-state initialization and update functions $(\phi_0,\phi)$. 

The agent state is an information state if it satisfies the following two properties:
\begin{enumerate}
    \item [\textbf{(P1)}] \textbf{Sufficient for performance evaluation.} There exists a function $r_{\IS} \colon \ALPHABET Z \times \ALPHABET A \to \reals$ such that for any $t$ and $H_t$ and $A_t$, we have
    \[
    \EXP[ R_t \mid H_t, A_t] = r_{\IS}( \vec \sigma_t(H_t), A_t).
    \]
    \item [\textbf{(P2)}] \textbf{Sufficient for predicting itself.} There exists a controlled transition probability matrix $P_{\IS} \colon \ALPHABET Z \times \ALPHABET A \to \Delta(\ALPHABET Z)$ such that for any $t$ and $H_t$ and $A_t$, we have
    \[
    \PR(Z_{t+1} \mid H_t, A_t) = P_{\IS}(Z_{t+1} \mid \vec \sigma_t(H_t), A_t).
    \]
\end{enumerate}

Recall that the agent state updates in a state-like manner~\eqref{eq:Z-update}. Thus,
\begin{multline*}
    \PR(Z_{t+1} \mid H_t, A_t) \\ = \sum_{y_{t+1} \in \ALPHABET Y}\PR(y_{t+1} \mid H_t, A_t)
    \IND_{\{Z_{t+1} = \phi(\vec \sigma_t(H_t), y_{t+1}, A_t)\}}
\end{multline*}
It was shown in~\cite{subramanian2022approximate} that the above relationship implies that a sufficient condition for (P2) is the following.
\begin{enumerate}
    \item [\textbf{(P2b)}] \textbf{Sufficient for predicting output.} There exists a controlled transition probability matrix $P_{\IS} \colon \ALPHABET Z \times \ALPHABET A \to \Delta(\ALPHABET Y)$ such that for any $t$ and $H_t$ and $A_t$, we have
    \[
    \PR(Y_{t+1} \mid H_t, A_t) = P_{\IS}(Y_{t+1} \mid \vec \sigma_t(H_t), A_t).
    \]
\end{enumerate}

The belief-state satisfies properties (P1) and (P2) and is, therefore, an information state. For specific models such as the linear quadratic Guassian (LQG) control or certain classes of machine repair models, simpler and model-specific information states exist. We refer the reader to~\cite{subramanian2022approximate} for a detailed discussion of the history of information states and various other examples.

A key implication of an information state is that it always leads to a dynamic programming decomposition.\footnote{We present multiple dynamic programming decompositions and differentiate between them via subscripts. A summary of these subscripts is shown in Table~\ref{tbl:DP}.} In particular, we have the following~\cite[Theorem 5 and 25]{subramanian2022approximate}. 
\begin{colortheorem}{Information-state based DP}{info-state}
    Suppose the agent state $\{Z_t\}_{t \ge 1}$ is an information state, i.e., satisfies properties (P1) and (P2) with some $(r_{\IS}, P_{\IS})$. Define the following dynamic program\footnote{For simplicity, we state the result for finite $\ALPHABET Z$. Similar decomposition holds for continuous $\ALPHABET Z$.}: 
    \begin{subequations}
    \begin{align}
        Q_{\IS}^\star(z,a) &= r_{\IS}(z,a) + \gamma \sum_{z' \in \ALPHABET Z} P_{\IS}(z'|z,a) V_{\IS}^\star(z'), \label{eq:Q}\\
        V_{\IS}^\star(z)   &= \max_{a \in \ALPHABET A} Q_{\IS}^\star(z,a). \label{eq:V}
    \end{align}
    Let $\pi_{\IS}^\star(z)$ denote any arg max of the right hand side of~\eqref{eq:V}. Then the policy $\boldsymbol{\vec \pi}_{\IS}^\star = (\vec \pi_{\IS,1}^\star, \vec \pi_{\IS,2}^\star, \dots)$ given by 
    \[
        \vec \pi_{\IS,t}^\star(h_t) = \pi_{\IS}^\star(\vec \sigma_t(h_t))
    \]
    is optimal, i.e., $J^{\boldsymbol{\vec \pi}^\star_{\IS}} = \JHND$.
    \end{subequations}
\end{colortheorem}
The classical belief-state based dynamic programming decomposition may be considered as an instance of \autoref{thm:info-state}. For specific models which have a simpler information state such as LQG control, \autoref{thm:info-state} provides a simpler dynamic program than the standard belief-state based dynamic program.

\begin{table}[!t]
    \centering
    \caption{Abbreviations used for different dynamic programs}
    \label{tbl:DP}
    \begin{tabular}{@{}cl@{}}
    \toprule
        \textbf{Abbreviation} & \multicolumn{1}{c}{\textbf{Meaning}}  \\
    \midrule
         \sublabel{IS} & Information-state based DP\\
         \sublabel{DES} & DP using designer's approach \\
         \sublabel{Prod} & DP based on product processes \\
         \sublabel{AIS} & Approximate information state based DP \\
         \sublabel{ASQL} & Agent-state based Q-learning \\
    \bottomrule
    \end{tabular}
\end{table}

In spite of its generality, \autoref{thm:info-state} is only applicable when the agent state satisfies properties (P1) and (P2), which is not always the case. A simple counterexample is when the agent uses a finite window of past observations with $Z_t = (Y_{t-n:t},A_{t-n:t-1})$. In the companion paper~\cite{serdar}, approximation results for this model are presented using ideas from filter stability. In the next section, we present other approaches that have been used in the literature for identifying optimal policies for models where properties (P1) and (P2) are not satisfied. 

\section{Optimal agent-state based policies}\label{sec:agent-state}
In this section, we consider the different policy classes (deterministic/stochastic and stationary/non-stationary) of agent-state based policies. We present examples to show that non-stationary policies may outperform stationary ones and that stochastic policies may outperform deterministic ones. We then present different schemes that have been used in the literature to identify optimal or sub-optimal policies within the different policy classes. In particular, the designer's approach for finding an optimal non-stationary agent-state based policy; policy evaluation and policy gradients for stationary agent-state based polices; and approximate information state approach for finding good stationary and deterministic agent-state based policies.
\subsection{Policy classes}

We start with a notation to denote (one-step) decision rules. Let
\begin{itemize}
    \item $\DRD = [\ALPHABET Z \to \ALPHABET A]$ denote the family of all deterministic decision rules, and
    \item $\DRS = [\ALPHABET Z \to \Delta(\ALPHABET A)]$ denote the family of all stochastic decision rules.
\end{itemize}

\begin{table}[!t]
    \centering
    \caption{List of different policy classes}
    \label{tbl:policy}
    \begin{tabular}{@{}cl@{}}
        \toprule
        \textbf{Policy class} & \multicolumn{1}{c}{\textbf{Meaning}}  \\
        \midrule
        $\PiHND$ & history based non-stationary deterministic policies \\ 
        $\PiHNS$ & history based non-stationary stochastic policies \\[5pt]
        
        $\PiZND$ & agent-state based non-stationary deterministic  policies \\ 
        $\PiZNS$ & agent-state based non-stationary stochastic  policies \\ 
        $\PiZSD$ & agent-state based stationary deterministic  policies \\ 
        $\PiZSS$ & agent-state based stationary stochastic  policies \\ 
         \bottomrule
    \end{tabular}
\end{table}

We define the following classes of agent-state based policies. Notation wise, we will use the absence of the vector accent to differentiate agent-state based policies from history-based policies. A summary of the different policy class used in this paper is shown in Table~\ref{tbl:policy}.
\begin{itemize}
    \item $\PiZND$ denotes the class of non-stationary agent-state based deterministic policies 
    $\boldsymbol{\pi} = (\pi_1, \pi_2, \dots)$ where $\pi_t \in \DRD$. 
    Let
    \[
        \JZND = \sup_{\boldsymbol{\pi} \in \PiZND} J^{\boldsymbol{\pi}}
    \]
    denote the optimal performance within class $\PiZND$. 
    A policy $\piZND \in \PiZND$ is called optimal in policy class $\PiZND$ if $J^{\piZND} = \JZND$.
    \item $\PiZNS$ denotes the class of non-stationary agent-state based stochastic policies 
    $\boldsymbol{\pi} = (\pi_1, \pi_2, \dots)$ where $\pi_t \in \DRS$.
    Let
    \[
        \JZNS = \sup_{\boldsymbol{\pi} \in \PiZNS} J^{\boldsymbol{\pi}}
    \]
    denote the optimal performance within class $\PiZNS$.
    A policy $\piZNS \in \PiZNS$ is called optimal in policy class $\PiZNS$ if $J^{\piZNS} = \JZNS$.
    \item $\PiZSD$ denotes the class of stationary agent-state based deterministic policies 
    $\boldsymbol{\pi} = (\pi, \pi, \dots)$ where $\pi \in \DRD$.
    Let
    \[
        \JZSD = \sup_{\boldsymbol{\pi} \in \PiZSD} J^{\boldsymbol{\pi}}
    \]
    denote the optimal performance within class $\PiZSD$.
    A policy $\piZSD \in \PiZSD$ is called optimal in policy class $\PiZSD$ if $J^{\piZSD} = \JZSD$.
    \item $\PiZSS$ denotes the class of stationary agent-state based stochastic policies 
    $\boldsymbol{\pi} = (\pi, \pi, \dots)$ where $\pi \in \DRS$.
    Let
    \[
        \JZSS = \sup_{\boldsymbol{\pi} \in \PiZSS} J^{\boldsymbol{\pi}}
    \]
    denote the optimal performance within class $\PiZSS$.
    A policy $\piZSS \in \PiZSS$ is called optimal in policy class $\PiZSS$ if $J^{\piZSS} = \JZSS$.
\end{itemize}
Note that $\PiZND$ and $\PiZNS$ agent-state based policies are different from $\PiHND$ and $\PiHNS$, which are history based policies. Therefore, $\JZND$ and $\JZNS$ are different from $\JHND$ and $\JHNS$. 

When the agent state is an information state and satisfies (P1) and (P2), \autoref{thm:info-state} implies that all the performance functions $\JZND$, $\JZNS$, $\JZSD$, $\JZSS$ are equal and are also equal to the performance of history-based policies $\JHND$ and $\JHNS$. 
However, when the agent state is not an information state, then the agent state does not satisfy the controlled Markov property, i.e., 
\[
    \PR(Z_{t+1} \mid Z_{1:t}, A_{1:t}) \neq \PR(Z_{t+1} \mid Z_t, A_t)
\]
or the property of being sufficient for reward evaluation, i.e., 
\[
    \EXP[r(S_t,A_t) \mid Z_{1:t}, A_{1:t}] \neq \EXP[r(S_t,A_t) \mid Z_t, A_t]
    .
\]
Moreover, the agent does not have perfect recall (i.e., at time~$t$, the agent does not have access to all the information that was available to the information that was available to it in the past). 

The absence of the information state properties imply that (F1)~non-stationary policies may outperform stationary policies and (F2)~stationary stochastic policies may outperform stationary deterministic policies.  These facts were first reported in~\cite{singh1994learning}. We illustrate them via examples borrowed from~\cite{Sinha2024}. 

\subsubsection*{(F1) Non-stationary agent-state based policies may outperform stationary agent-state based policies}
\begin{figure}[!t]
    \centering
    \includegraphics[width=0.9\linewidth]{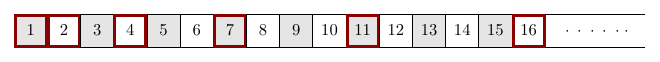}
    \caption{The cells indicate the state of the environment. Cells with the same background color have the same observation. The cells with a thick red boundary correspond to elements of the set $\ALPHABET D_0 \coloneqq \{ n(n+1)/2 + 1 : n \in \naturalnumbers \}$, where the action~$0$ gives a reward of $+1$ and moves the state to the right, while the action~$1$ gives a reward of $-1$ and resets the state to~$1$. The cells with a thin black boundary correspond to elements of the set $\ALPHABET D_1 = \naturalnumbers \setminus \ALPHABET D_0$, where the action~$1$ gives the reward of $+1$ and moves the state to the right while the action~$0$ gives a reward of $-1$ and resets the state to~$1$.}
    \label{fig:ex1}
\end{figure}

Consider the POMDP described in \autoref{fig:ex1}, where the system starts in state~$1$.

Since the dynamics are deterministic,  the agent can infer the current state from the history of past actions and can take the action to increment the current state and receive a per-step reward of $+1$. Thus, $\JHND = 1/(1-\gamma)$. Furthermore, since the system is deterministic, the optimal policy can be implemented via an open-loop policy given by $\{a^\star_t\}_{t \ge 1}$, where $a^\star_t = \IND_{\{ t \in \ALPHABET D_1 \}}$, which can be implemented irrespective of the information structure. Thus, $\JZND = \JHND$. 

When $Z_t = Y_t$, $\PiZSD$ consists of four policies. A brute force evaluation shows that $\JZSD = (1 + \gamma - \gamma^2)/(1 - \gamma^3) < \JZND$. See~\cite{Sinha2023} for details. Thus, non-stationary agent-state policies may outperform stationary agent-state based policies.\footnote{The conclusion continue to hold if we compare with the performance on stationary stochastic policies. Although it is not possible to evaluate $\JZSS$ in closed form, a brute force Monte Carlo evaluation shows that $\JZSS \approx \JZSD$. Therefore, $\JZSS < \JZND$.}

\subsubsection*{(F2) Stationary stochastic agent-state policies may outperform stationary deterministic agent-state based policies}
\begin{figure}
    \centering
    \begin{subfigure}{0.45\linewidth}
    \centering
    \scalebox{0.7}
    {\begin{tikzpicture}[line width=1pt, node distance=2cm and 2cm]
        \useasboundingbox (-0.5,-0.65) rectangle (4.5,1.25);
        \node[state] (M) {$0$};
        \node[state, right of=M] (Z) {$1$};
        \node[state, right of=Z] (P) {$2$};

        \draw[every loop]
            (Z) edge[bend right, auto=right] node {$0.5$} (M)
            (Z) edge[bend left, auto=left] node {$0.5$} (P)
            (M) edge[loop above] node {$1$} (M)
            (P) edge[loop above] node {$1$} (P)
            ;

    \end{tikzpicture}}
    \caption{Dynamics under action~$0$}
    \end{subfigure}
    \hfill
    \begin{subfigure}{0.45\linewidth}
    \centering
    \scalebox{0.7}
    {\begin{tikzpicture}[line width=1pt, node distance=2cm and 2cm]
        \useasboundingbox (-0.5,-0.65) rectangle (4.5,1.25);
        \node[state] (M) {$0$};
        \node[state, right of=M] (Z) {$1$};
        \node[state, right of=Z] (P) {$2$};

        \draw[every loop]
            (M) edge[bend right, auto=right] node {$0.5$} (Z)
            (P) edge[bend left, auto=left] node {$0.5$} (Z)
            (M) edge[loop above] node {$0.5$} (M)
            (P) edge[loop above] node {$0.5$} (P)
            (Z) edge[loop above] node {$1$} (Z)
            ;
            
    \end{tikzpicture}}
    \caption{Dynamics under action~$1$}
    \end{subfigure}
    \caption{A POMDP with $\ALPHABET S = \{0,1,2\}$, $\ALPHABET A = \{0, 1\}$ and $\ALPHABET Y = \{0\}$. The rewards functions are $r(\cdot,0) = [-1,0,2]$ and $r(\cdot, 1) = -0.5$.}
    \label{fig:ex2}
    \includegraphics[width=0.9\linewidth]{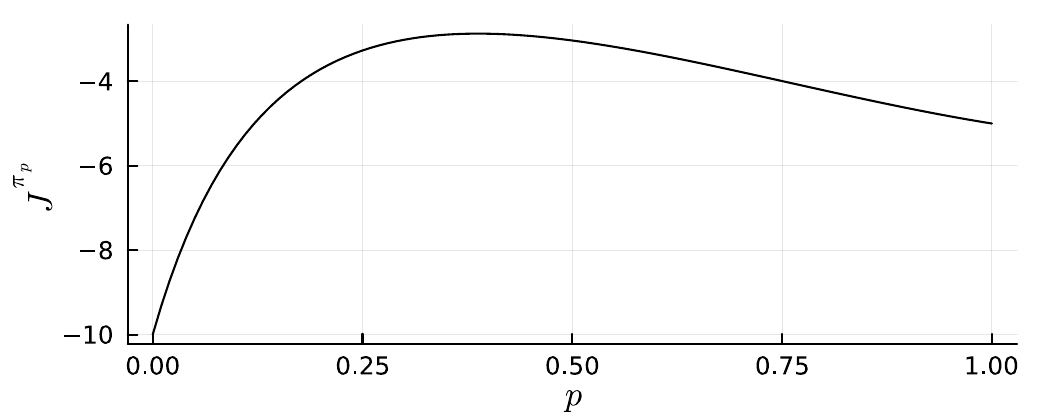}
    \caption{Performance of stationary stochastic policies in the model of \autoref{fig:ex2}.}
    \label{fig:ex2-perf}
\end{figure}

Consider the POMDP shown in \autoref{fig:ex2}, where the system starts in state $0$. The agent gets no observations, i.e., $Y_t \equiv 0$, and we consider agent-state policies with $Z_t = Y_t \equiv 0$. A policy $\boldsymbol\pi \in \PiZSS$ is parameterized by a single parameter $p \in [0,1]$, which indicates the probability of choosing action~$1$. We denote such a policy by $\boldsymbol\pi_p$. Note that if $p \in \{0,1\}$, then $\boldsymbol\pi_p \in \PiZSD$. Let $(P_a, r_a)$ denote the probability transition matrix and reward function when $a \in \ALPHABET A$ is chosen and let $(P_p, r_p) = (1-p)(P_0, r_0) + p(P_1, r_1)$. Then, the performance of policy $\boldsymbol\pi_p$ is given by $J^{\boldsymbol\pi_p} = [(1 - \gamma P_p)^{-1} r_p]_{0}$. The performance for all $p \in [0,1]$ for $\gamma = 0.9$ is shown in \autoref{fig:ex2-perf}, which shows that the best performance is achieved by the stochastic policy $\boldsymbol\pi_p$ with $p \approx 0.39$.

In summary, \bolditalic{the optimal agent-state based policy depends on the choice of the policy class}. The relationship between the different performance measures can be summarized as follows:
\begin{tcolorbox}[ams equation, colorformula]
\label{eq:diagram}
    \begin{tikzcd}
        \JZSD \arrow{d} \arrow{r} & \JZND \arrow[xshift=0.25em]{d} \arrow{r} & \JHND \arrow[xshift=0.25em]{d} \\
        \JZSS \arrow{r} & \JZNS \arrow{r} \arrow[xshift=-0.25em,dashed]{u}& \JHNS \arrow[xshift=-0.25em,dashed]{u} 
    \end{tikzcd}
\end{tcolorbox}
\noindent
where the arrows mean ``less than or equal to''. The solid arrows follow from set inclusion relationships (that is, one policy class is a subset of the other); the dashed arrow relationships need to established explicitly. The reason for the dashed arrow between $\JHNS$ and $\JHND$ has already been presented in the discussion around~\eqref{eq:history}. The reason for the dashed arrow between $\JZNS$ and $\JZND$ will be presented later in~\eqref{eq:non-stationary}.

Features (F1)--(F2) may appear to be surprising because they are not present in MDPs or POMDPs when using a belief state. In fact, they are absent when the agent state is an information state. However, when the agent state is not an information state, the system has what is called a \emph{non-classical information structure}\footnote{See~\cite{Mahajan2012} for a general discussion of non-classical information structures.} and, therefore, the problem of finding the optimal agent-state based policy is a \emph{decentralized} stochastic control problem (even though there is only one decision maker). This fact is well known in the decentralized control literature~\cite{Sandell1974,MahajanPhD} but perhaps not as well recognized in the POMDP literature.

The fact that the optimization problem at hand is a decentralized control problem means that we cannot directly use the standard results from Markov decision theory to find the optimal policy in a given policy class. In the remainder of this section, we summarize the approaches that have been taken in the literature to find the optimal or a good policy within a policy class. 

\subsection{The designer's approach to find the optimal non-stationary policy}

Since the problem of finding the optimal non-stationary agent-state policy is a decentralized stochastic control policy, it is possible to use solution techniques from decentralized stochastic control to find the optimal policy in class $\PiZND$ and $\PiZNS$. One such approach is the designer's approach, which was proposed in~\cite{MahajanPhD} and can be viewed as a refinement of an earlier approach known as the standard form proposed in~\cite{Witsenhausen1973}. The designer's approach was presented in~\cite{MahajanPhD} for a two-agent decentralized control system where each agent followed an agent-state based policy. We present a simplified version of this approach, restricted to the single agent POMDP setting.

For any policy $\boldsymbol{\pi} = (\pi_1, \pi_2, \dots) \in \PiZNS$, define $\xi^{\boldsymbol\pi}_t \in \XIS$ by
\[
    \xi^{\boldsymbol{\pi}}_t(s,z) = \PR^{\boldsymbol{\pi}}(S_t = s, Z_t = z),
    \quad \forall s \in \ALPHABET S, z \in \ALPHABET Z.
\]
Since $\PiZSD, \PiZSS, \PiZND \subseteq \PiZNS$, the same definition holds for any policy in $\PiZSD, \PiZSS, \PiZND$ as well. The key idea of the designer's approach is to show that the process $\{\xi^{\boldsymbol{\pi}}\}_{t \ge 1}$ is a controlled Markov process controlled by $\{\pi_t\}_{t \ge 1}$.

We first start with some definitions.
\begin{itemize}
    \item 
        Define a function $\phi_{\Des} \colon \XIS \times \DRS \to \XIS$ as follows: for any $\xi \in \XIS$, $\pi \in \DRS$ and for all $s',z' \in \ALPHABET S \times \ALPHABET Z$, 
        \begin{align*}
        &[\phi_{\Des}(\xi,\pi)](s',z') 
        \\
        &=
            \smashoperator{\sum_{\substack{(s,z,y',a) \in \\ \ALPHABET S \times \ALPHABET Z \times \ALPHABET Y \times \ALPHABET A}}}
            \xi(s,z) \pi(a\,|\,z) P(y',s'|s,a) \IND_{\{ z' = \phi(z,y',a) \}}
        \end{align*}
    \item 
        Define the function $r_{\Des} \colon \XIS \times \DRS \to \reals$ as follows: for any $\xi \in \XIS$ and $\pi \in \DRS$:
        \[
            r_{\Des}(\xi, \pi) \coloneqq 
            \smashoperator{\sum_{\substack{(s,z,a) \in \\ \ALPHABET S \times \ALPHABET Z \times \ALPHABET A}}}
            \xi(s,z) \pi(a \,|\, z) r(s,a).
        \]
\end{itemize}
Note that both $\phi_{\Des}$ and $r_{\Des}$ are bilinear functions.%
\footnote{Since $\XIS$ and $\DRS$ are bounded sets, linearity in the each component should be interpreted as follows:
for any $\xi^1, \xi^2 \in \XIS$, $\pi \in \DRS$, and $\lambda \in [0,1]$, we have
    \[
        \phi_{\Des}(\lambda \xi^1 + (1-\lambda) \xi^2, \pi)
        =
        \lambda \phi_{\Des}(\xi^1,\pi) + (1-\lambda) \phi_{\Des}(\xi^2,\pi)
    \]
    with a similar interpretation for $r_{\Des}$ and the $\pi$ component of both functions.}

Then a simple application of Bayes rule and the definition of expectation imply the following~\cite{MahajanPhD}.
\begin{proposition}\label{prop:designer}
    The process $\{\xi^{\pi}_t\}_{t \ge 1}$ is a controlled Markov process controlled by $\{\pi_t\}_{t \ge 1}$, i.e.,
    \begin{enumerate}
        \item $\xi^{\boldsymbol{\pi}}_{t+1} = \phi_{\Des}(\pi_t, \xi^{\boldsymbol{\pi}}_t)$;
        \vskip 0.25\baselineskip
        \item $\EXP^{\boldsymbol{\pi}}[ r(S,A) ] = r_{\Des}(\pi_t, \xi^{\boldsymbol{\pi}}_t)$.
    \end{enumerate}
\end{proposition}

The designer's approach is based on the following interpretation. Consider the system designer who wants to pick a policy $\boldsymbol{\pi} \in \PiZNS$ (or $\PiZND$). From the point of view of such a designer, the system is a completely unobserved input-output system, where the system designer chooses the control input $\pi_t$, receives a per-step reward $r(S_t,A_t)$, but does not observe anything. \autoref{prop:designer} implies that $\xi^{\boldsymbol{\pi}}_t$ is an information state for this system. Therefore, the system designer can use dynamic programming to identify its ``control actions''. Consequently, we have the following result~\cite{MahajanPhD}.

\begin{colortheorem}{DP using the designer's approach}{designer}
    Consider the following dynamic program: for all $\xi \in \XIS$,
    \begin{equation}
        V_{\Des}(\xi) 
        = 
        \max_{ \pi \in \DRS}
        \bigl\{
        r_{\Des}(\xi,\pi)
        + 
        \gamma V_{\Des}(\phi_{\Des}(\xi, \pi))
        \bigr\}.
        \label{eq:designer-DP}
    \end{equation}
    Let $\psi_{\Des}(\xi)$ denote any arg max of the right hand side of~\eqref{eq:designer-DP}. 
    Let $\xi_1(s_1,a_1) \coloneqq \PR(S_1 = s_1, Z_1 = z_1)$ denote the initial distribution of the system and the agent state. Recursively define $\{\xi^\star_t\}_{t\ge 1}$ and $\{\pi^\star_t\}_{t \ge 1}$ as follows: $\xi^\star_1 = \xi_1$ and for $t \ge 1$:
    \[
        \pi^\star_t = \psi_{\Des}(\xi^\star_t)
        \quad\text{and}\quad
        \xi^\star_{t+1} = \phi_{\Des}(\xi^\star_t, \pi^\star_t).
    \]
    Then the policy $\boldsymbol{\pi}^\star = (\pi^\star_1, \pi^\star_2, \dots) \in \PiZNS$ is the optimal policy in $\PiZNS$. 
\end{colortheorem}

\subsection*{Some remarks on the designer's approach}
\begin{enumerate}
    \item The designer's approach presented here is adapted from the presentation in~\cite{MahajanPhD} for decentralized control problems with two agents, where the map $\psi_{\Des} \colon \XIS \to \DRS$ was called a \bolditalic{meta-policy}. The general idea of using the probability distribution over all pertinent variables as an information state goes back to~\cite{Witsenhausen1973}. The method was re-discovered in~\cite{Dibangoye2016}, where it was called occupation MDP. A similar idea was presented in~\cite{Bagnell2003} for MDPs and POMDPs with memoryless policies.
    \item In the argument above, we maximized over $\pi \in \DRS$ in~\eqref{eq:designer-DP}. If we instead maximize over $\pi \in \DRD$, we will obtain the optimal policy $\boldsymbol{\pi}^\star \in \PiZND$. 
    \item The probability distribution $\xi^{\boldsymbol{\pi}}_t$ may be viewed as the ``belief'' of the designer on the state sufficient for input-output mapping (i.e., $(S_t,Z_t)$) given the history of past observations and actions (i.e., $\pi_{1:t-1}$, since the designer does not observe anything). Therefore, as stated in~\cite{MahajanPhD} and~\cite{Dibangoye2016}, we can follow the standard argument for POMDPs~\cite{Smallwood1973} to argue that the value function $V_{\Des}$ is convex (it is piecewise linear and convex for finite horizon models).
    \item Since $V_{\Des}$ is convex and $\phi_{\Des}$ and $r_{\Des}$ are bilinear, it means that 
    \[
        Q_{\Des}(\xi,\pi) \coloneqq r_{\Des}(\xi,\pi) + \gamma V_{\Des}(\phi_{\Des}(\xi, \pi))
    \]
    is convex in $\pi$. Therefore, for any fixed $\xi$, the maximum value of the convex function $Q_{\Des}(\xi, \pi)$ over $\pi \in \DRS$ (which is a convex polyhedron) is achieved at a vertex of $\DRS$. Thus, the arg max in~\eqref{eq:designer-DP} is a deterministic decision rule $\pi \in \DRD$. 
    Therefore,
    \begin{equation}\label{eq:non-stationary}
        \tcbhighmath{\JZND = \JZNS.}
    \end{equation}
    Consequently, \bolditalic{there is no loss of optimality in restricting attention to non-stationary deterministic (rather than stochastic) agent-state based policies.}
    \item As far as we are aware, the result in~\eqref{eq:non-stationary} is new. A similar result is claimed in~\cite[Proposition 1]{Bagnell2003} for memoryless policies in POMDPs (i.e., $Z_t = Y_t$), but for a different definition of optimality. 
    \item Note that the dynamics of $\{\xi_t\}_{t \ge 1}$ in~\eqref{eq:designer-DP} are deterministic. Therefore, it is possible to use ideas from deterministic optimization to find the optimal \emph{trajectory} $\{\xi^\star_t\}_{t \ge 1}$ and optimal \emph{open-loop} ``control actions'' $\{\pi^\star_t\}_{t \ge 1}$. 
\end{enumerate}

\subsection{Policy search methods to find a locally optimal policy in $\PiZSS$}\label{sec:product}

We can use standard results from policy search methods for MDPs to identify a stationary policy that is locally optimal (within the class of stationary policies). The high-level idea is based on the fact that
the \emph{product process} $\{(S_t,Z_t)\}_{t \ge 1}$ is a controlled Markov process with the controlled transition probability:
    \begin{align*}
        &P_{\PROD}(s',z' \mid s,z, a) 
        \\
        &\quad \coloneqq 
        \sum_{y' \in \ALPHABET Y}
        P(s',y' \mid s,a) \IND_{\{ z' = \phi(z,y',a)\}}.
    \end{align*}
    
    Now, for any $\boldsymbol{\pi} = (\pi, \pi, \dots) \in \PiZSS$, define:
    \[
        P^{\boldsymbol\pi}_{\PROD}(s',z' \mid s,z) 
        \coloneqq \sum_{a \in \ALPHABET A} 
        \pi(a|z) P_{\PROD}(s',z'|s,z,a).
    \]
    and
    \[
        r^{\boldsymbol\pi}_{\PROD}(s,z) \coloneqq \sum_{a \in \ALPHABET A} \pi(a|z) r(s,a).
    \]
    Let $V^{\boldsymbol\pi}_{\PROD} \colon \ALPHABET S \times \ALPHABET Z \to \reals$ be the value function of the policy $\boldsymbol\pi$ in the $(S,Z)$ space, i.e., it is the solution of the following policy evaluation formula:
    \begin{align}\label{eq:policy-evaluation}
        V^{\boldsymbol\pi}_{\PROD}(s,z) &= r^{\boldsymbol\pi}_{\PROD}(s,z)  \nonumber \\
        &\quad +
        \gamma \smashoperator{\sum_{\substack{(s',z') \in \\ \ALPHABET S \times \ALPHABET Z}}}
        P^{\boldsymbol\pi}_{\PROD}(s',z'|s,z) V^{\boldsymbol\pi}_{\PROD}(s',z'). 
    \end{align}
    Let $Q^{\boldsymbol\pi}_{\PROD} \colon \ALPHABET S \times \ALPHABET Z \times \ALPHABET A \to \reals$ denote the action-value function of the policy $\boldsymbol\pi$, i.e., for any $(s,z,a) \in \ALPHABET S \times \ALPHABET Z \times \ALPHABET A$, 
    \begin{align*}
        Q^{\boldsymbol\pi}_{\PROD}(s,z,a) 
        &=
        r(s,a) \\
        &\quad + \gamma \smashoperator{\sum_{\substack{(s',z') \in \\ \ALPHABET S \times \ALPHABET Z}}}
        P_{\PROD}(s',z'|s,z,a) V^{\boldsymbol\pi}_{\PROD}(s',z').
    \end{align*}
    Finally, let $d^{\boldsymbol{\pi}}_{\PROD}$ denote the \emph{unnormalized} occupancy measure of the policy $\boldsymbol{\pi}$, i.e., 
    \[
        d^{\boldsymbol\pi}_{\PROD}(s,z,a) = \sum_{t=1}^\infty \gamma^{t-1} \PR^{\boldsymbol\pi}(S_t = s, Z_t = z, A_t = a).
    \]
    Then, the performance of policy $\boldsymbol\pi \in \PiZSS$ is given by
    \[
        J^{\boldsymbol{\pi}} = \sum_{(s,z) \in  \ALPHABET S \times \ALPHABET Z}
        \xi_1(s,z) V^{\boldsymbol\pi}_{\PROD}(s,z),
    \]
    where $\xi_1(s,z) \coloneqq \PR(S_1 = s, Z_1 = z)$ denotes the initial distribution of the joint state of the environment and the agent.  This shows that performance evaluation of any policy in $\PiZSS$ is straightforward.

    Given the above performance evaluation formula, we can use policy-gradient based methods to find a good policy. In particular, 
     suppose $\boldsymbol\pi = (\pi,\pi,\dots) \in \PiZSS$ is a parameterized policy (e.g., softmax or mixture of Gaussians) with policy parameters $\theta$. We will use $\boldsymbol\pi_\theta$ to denote the policy with parameters $\theta$ and $\pi_\theta$ to denote the decision rule with parameters~$\theta$. 
    Then, by a straight forward modification of the policy gradient formula~\cite{Sutton1999,Konda1999,Konda2003,Cao2007}, we get
    \begin{tcolorbox}[ams equation, colorformula]
        \hskip -1em
        \GRAD_{\theta} J^{\boldsymbol\pi_{\theta}}
        =
        \smash{\smashoperator{\sum_{\substack{(s,z,a) \in \\ \ALPHABET S \times \ALPHABET Z \times \ALPHABET A}}}}
        \vphantom{\bigg(}
        d^{\boldsymbol\pi_\theta}_{\PROD}(s,z,a) Q^{\boldsymbol\pi_\theta}_{\PROD}(s,z,a) 
        \GRAD_{\theta} \log \pi_{\theta}(a | z).
        \label{eq:policy-gradient}
    \end{tcolorbox}

Consequently, we can use any policy gradient based search algorithm (such as actor-critic and its variants; see~\cite{Weng2018PG}) to find locally optimal policy parameters, i.e., policy parameters $\theta^\star$ such that 
\[
    \GRAD_{\theta} J^{\boldsymbol\pi_{\theta}} \Big|_{\theta = \theta^\star} = 0.
\]
Then the policy $\boldsymbol\pi_{\theta^\star}$ is a locally optimal policy in class $\PiZSS$. 

\subsection*{Some remarks on the policy gradient approach}
\begin{enumerate}
    \item The policy evaluation formula~\eqref{eq:policy-evaluation} using the product state $(S_t, Z_t)$ was initially presented in~\cite{Platzman1977} and has been rediscovered in slightly different forms multiple times~\cite{Littman1996,Cassandra1998,Hauskrecht1997,Hansen1998}.
    
    \item There is a rich history of policy  based algorithms in Systems and Control, Operations Research, and Artificial Intelligence. We refer the reader to~\cite{Baxter2001,Cao2003} for detailed overviews.

    \item Several other approaches for policy search in $\PiZNS$ have been proposed in the literature~\cite{Hansen1998,PEGASUS,krishnamurthy2016partially,kochenderfer2022algorithms}.
    
    \item Note that the formula in~\eqref{eq:policy-gradient} is applicable when the system dynamics are known (and therefore the $Q^{\boldsymbol{\pi}}_{\PROD}$ can be computed via policy evaluation formula described above). In the RL setting, the environment state $S_t$ is not observed, so the above formula cannot be used directly. 
    
    \item For RL, the idea of using policy gradient methods for POMDPs was first presented in~\cite{Kimura1997} and variations have been proposed in~\cite{Jaakkola1994,Briad1998,Meuleau1999,Peshkin1999,Aberdeen2002,Yu2006,krishnamurthy2016partially}.
    However, these policy gradient formula proposed in these papers is for the RL setting where the agent state $S_t$ is not observed; therefore the exact expressions were different. We present an alternative approach in \autoref{sec:RL}.
    
    \item  As pointed out in~\cite{Kimura1997}, the agent state in a POMDP can be viewed as a \emph{feature} of the entire history of observations and actions. With this viewpoint, the policy gradient formula~\eqref{eq:policy-gradient} may be viewed as a special case of actor-critic algorithms with features~\cite{Konda1999,Konda2003}.
    \item Another idea that has been used in the RL setting is asymmetric actor critic~\cite{Baisero2022,Sinha2023}, where it is assumed that the critic has access to the environment state (which is the case in simulation environments). In such settings, it is possible to learn $Q^{\boldsymbol\pi}(s,z,a)$ via temporal difference learning. See~\cite{Baisero2022,Sinha2023} for a discussion.
\end{enumerate}

\subsection{The approximate information state approach to find a good policy in $\PiZSD$}\label{sec:ais}

In this section, we present an alternative approach to finding good policies in $\PiZSD$ called the \emph{approximate information state} (AIS), which was originally developed in~\cite{subramanian2022approximate}. We first start with the intuition and then explain the technical results.

\subsubsection*{Intuition}
One way to obtain a policy is to posit \emph{any} dynamics and rewards $(P_{\AIS}, r_{\AIS})$, where $P_{\AIS} \colon \ALPHABET Z \times \ALPHABET A \to \Delta(\ALPHABET S)$ and $r_{\AIS} \colon \ALPHABET Z \times \ALPHABET A \to \reals$, and obtain a policy $\boldsymbol{\vec \pi}_{\AIS}$ by solving the following dynamic program:
\begin{subequations}\label{eq:ais-DP}
\begin{align}
    Q_{\AIS}(z,a) &= r_{\AIS}(z,a) + \gamma \sum_{z' \in \ALPHABET Z} P_{\AIS}(z'|z,a) V_{\AIS}(z'), \label{eq:QAIS}\\
    V_{\AIS}(z)   &= \max_{a \in \ALPHABET A} Q_{\AIS}(z,a). \label{eq:VAIS}
\end{align}
\end{subequations}
Let $\pi_{\AIS}(z)$ denote any arg max of the right hand side of~\eqref{eq:VAIS}. Define the policy $\boldsymbol{\vec \pi}_{\AIS} = (\vec \pi_{\AIS,1}, \vec \pi_{\AIS,2}, \dots)$ given by\footnote{Recall that $\boldsymbol{\vec \sigma} = (\vec \sigma_1, \vec \sigma_2, \dots)$ is the history compression function corresponding to the agent-state update function $\phi$ and is given by~\eqref{eq:history-compression}.}
\begin{equation}\label{eq:ais-policy}
    \vec \pi_{\AIS,t}(h_t) = \pi_{\AIS}(\vec \sigma_t(h_t))
\end{equation}

As shown in \autoref{thm:info-state}, if the posited dynamics and rewards $(P_{\AIS}, r_{\AIS})$ satisfy properties (P1) and (P2), the policy $\boldsymbol{\vec \pi}_{\AIS}$ is optimal. But what happens when properties (P1) and (P2) are not satisfied? 

Let $(\boldsymbol \varepsilon, \boldsymbol \delta)$ with $\boldsymbol \varepsilon = (\varepsilon_1, \varepsilon_2, \dots)$ and $\boldsymbol \delta = (\delta_1, \delta_2, \dots)$ with $\varepsilon_t, \delta_t \in \reals_{\ge 0}$ be such that the following properties are satisfied:
\begin{enumerate}
    \item [\textbf{(AP1)}] \textbf{Approximately sufficient for performance evaluation.} For any~$t$, and $H_t$ and $A_t$ we have
    \[
    \bigl| \EXP[ R_t \mid H_t, A_t] - r_{\AIS} (\vec \sigma_t(H_t), A_t) \bigr| \le \varepsilon_t.
    \]
    \item [\textbf{(AP2)}] \textbf{Sufficient for predicting itself.} For any~$t$, and $H_t$ and $A_t$ we have
    \[
    d_{\F}(\PR(Z_{t+1} \mid H_t, A_t), \space P_{\AIS}(Z_{t+1} \mid \vec \sigma_t(H_t), A_t)) \le \delta_t.
    \]
    where $d_{\F}$ is a metric on $\Delta(\ALPHABET Z)$, which we will make precise later. 
\end{enumerate}
Then the tuple $(\boldsymbol{\vec \sigma}, P_{\AIS}, r_{\AIS})$ is called an \emph{approximate information state} (AIS). 

Engineering intuition suggests that the sub-optimality of $\boldsymbol{\vec \pi}_{\AIS}$, i.e., 
\[
   \alpha \coloneqq \JHND - J^{\boldsymbol{\vec \pi}_{\AIS}}
\]
should be a continuous function of $(\boldsymbol \varepsilon, \boldsymbol \delta)$. The results of~\cite{subramanian2022approximate} formalize this intuition.

\subsubsection*{The choice of metric}
To formalize the above intuition, we need to choose a metric $d_{\F}$ on probability spaces. For our purposes, it is convenient to work with a class of metrics known as integral probability measures (IPMs)~\cite{muller1997integral}. 
Let $\ALPHABET V$ denote the family of all bounded (measurable) functions from $\ALPHABET Z$ to $\reals$ and $\ALPHABET P$ denote the set of all probability measures on $\ALPHABET Z$ with finite mean.\footnote{When $\ALPHABET Z$ is finite $\ALPHABET V \equiv \reals^{|\ALPHABET Z|}$ and $\ALPHABET P$ is the set of all probability mass functions on $\ALPHABET Z$.}

\begin{definition}\label{def:IPM}
    Let $\F$ be a convex and balanced\footnote{A subset $\F$ of $\ALPHABET V$ is balanced if for all $f \in \F$ and scalars $a$ such that $|a| \le 1$, $a f \in \F$.} subset of $\ALPHABET V$. Then, 
    the IPM distance (w.r.t.\ $\F$) between two probability laws $\nu_1, \nu_2 \in \ALPHABET P$ is given by 
    \[
        d_{\F}(\nu_1, \nu_2) = \sup_{f \in \mathfrak {F}}\biggr|\int f d\nu_1 - \int f d\nu_2 \biggl|.
    \]
\end{definition}
\begin{definition}
    In the setting of Definition~\ref{def:IPM}, the  Minkowski functional of any measurable function $f \in \ALPHABET V$ is defined as
    \[
        \rho_{\F}(f) = \inf\Bigl\{\rho \in \mathbb R_{>0}: \frac{f}{\rho}\in \F\Bigr\}.
    \]
\end{definition}
Note that if for every positive $\rho$, $f/\rho \not\in \F$, then $\rho_{\F}(f) = \infty$.

An immediate consequence of the above two definitions is that for any measurable function $f \in \ALPHABET V$,
\begin{equation}\label{eq:IPM_main}
    \biggl| \int f d\nu_1 - \int f d\nu_2 \biggr| \leq \rho_{\F}(f)d_{\F}(\nu_1,\nu_2).
\end{equation}
Many of the commonly used metrics on probability spaces are IPMs. For example

\begin{itemize}
    \item \textbf{Total variation distance}, denoted by $\dTV$, corresponds to $\F = \FTV \coloneqq \{ f \in \ALPHABET V : \tfrac 12\Span(f) \le 1 \}$, where $\Span(f) = \sup(f) - \inf(f)$~\cite{muller1997integral,villani2009optimal}. In this case $\rho_{\F}(f) = \tfrac 12 \Span(f)$.
    \item \textbf{Wasserstein distance}. Suppose $(\ALPHABET Z, d_{\ALPHABET Z})$ is a metric space. Then the Wasserstein distance, denoted by $\dWas$, corresponds to $\F = \FWas \coloneqq \{ f \in \ALPHABET V : \Lip(f) \le 1 \}$ where  $\Lip(f)$ denotes the Lipschitz constant of a function $f$~\cite{vaserstein1969markov,villani2009optimal}. In this case $\rho_{\F}(f) = \Lip(f)$.
\end{itemize}
Other distances such as Kantorovich distance, bounded Lipschitz, maximum mean discrepancy, are also instances of IPM. See~\cite{muller1997integral,subramanian2022approximate} for details.

We now state the main result of~\cite{subramanian2022approximate}.
\begin{colortheorem}{AIS based DP}{ais}
    Given a function class $\F$, let $(\boldsymbol{\vec \sigma}, P_{\AIS}, r_{\AIS})$ be an $(\boldsymbol \varepsilon, \boldsymbol \delta)$-AIS with respect to $d_{\F}$. Consider the DP~\eqref{eq:ais-DP}. Then, the policy $\boldsymbol{\vec \pi}_{\AIS}$ given by~\eqref{eq:ais-policy} satisfies
    \begin{equation}
        \JHND - J^{\boldsymbol{\vec \pi}_{\AIS}} \le 
        \frac{2}{1-\gamma} \bigl[ \varepsilon + \gamma \delta \rho_{\F}(V^{\star}_{\AIS}) \bigr],
    \end{equation}
    where
    \begin{align*}
        \varepsilon &= (1-\gamma) \sum_{t=1}^\infty \gamma^{t-1} \varepsilon_t, &
        \delta &= (1-\gamma) \sum_{t=1}^\infty \gamma^{t-1} \delta_t.
    \end{align*}
\end{colortheorem}

\subsection*{Some remarks on AIS}
\begin{enumerate}
    \item We have presented the simplest form of the AIS result. Several variations including finite horizon models, multi-agent systems, and others are presented in~\cite{subramanian2022approximate}. 
    \item The results of \autoref{thm:ais} may be viewed as a generalization of approximation bounds for MDPs. The earliest such result is~\cite{muller1997}. Several variations of such results for MDPs have been derived in the literature, including model approximation in MDPs~\cite{Asadi2018}, state abstraction in MDPs~\cite{Li2006,Asadi2018,gelada2019deepmdp}. As shown in~\cite{subramanian2022approximate}, many of these results are special instances of \autoref{thm:ais}.
    \item When defining information state, it was mentioned that (P2b) is a sufficient condition for (P2). Along similar lines, condition (AP2) can be relaxed to approximately predicting the next observation. See~\cite{subramanian2022approximate} for details.
    \item One option for solving belief-state based POMDPs is by discretizing the belief space. Approximation bounds for belief discretization were derived in~\cite{FrancoisLavet2019} and it was shown in~\cite{subramanian2022approximate} that such bounds may be viewed as a special instance of \autoref{thm:ais}. 
    \item A heuristic algorithm to jointly learn a good AIS representation and a agent-state based policy is proposed in~\cite{yang2022discrete}.
    \item In addition to the AIS approach presented here, there are two popular approaches to determine good policies in $\PiZSD$: bisimulation metrics~\cite{CastroPanangadenPrecup_2009,castro2021mico} and predictive state representations~\cite{RosencrantzGordonThrun_2004,Boots2011,Hamilton2014,Kulesza2015,Kulesza2015a,Jiang2016}. These approaches are based on different approximation philosophies and are conceptually different.
    \item The results of \autoref{thm:ais} are also applicable for worst case design. In particular, let the \emph{worst-case distance} $d_{\F}(X_1, X_2)$ between two bounded random variables $X_1$ and $X_2$ be defined as the Hausdorff distance between their supports. It is then shown in~\cite{dave2022worst,dave2023worst,dave2024worst} that under an additional smoothness assumption on the dynamics, one may define an AIS in terms of the the worst-case distance and generalize the results of \autoref{thm:ais} to provide worst case guarantees for agent-state based policies.
\end{enumerate}

\section{Reinforcement learning approaches to learn a policy in $\PiZSD$ or $\PiZSS$} \label{sec:RL}

As highlighted in the introduction, one of the motivations for considering agent-state based policies is that they are easier to use in the RL setting because the agent-state update does not depend on the system model. In this section, we review the different approaches for RL in POMDPs with agent state. 

Before stating the results, we start with a remark that is not sufficiently highlighted in the literature. In the standard POMDP model, it is assumed that the actions are a function of the history of past observations and actions. In the learning setup, actions are also sometimes allowed to be a function of the history of rewards. Therefore, in the literature on RL for POMDPs, it is always implicitly assumed  that the observations include the rewards as well. If that is not the case, we need to consider an agent-state update rule of the form $Z_{t+1} = \phi(Z_t, Y_{t+1}, A_t, R_{t+1})$, but that does not fundamentally change the nature of the results. 

\subsection{Agent-state based Q-learning (ASQL)}

One of the simplest RL algorithms is Q-learning~\cite{watkins1992QL}, where the agent learns the Q-function by bootstrapping. For MDPs, it is well understood that the Q-learning iterates converge to the optimal Q-function, which is the solution of the dynamic programming equation. 

Agent-state based Q-learning (ASQL) uses the same idea for POMDPs with agent state. In the simplest setting, the agent acts in the environment according to a behavior policy $\mu \colon \ALPHABET Z \to \Delta(\ALPHABET A)$ and generates the experience $(Y_1, Z_1, A_1, R_1, Y_2, Z_2, A_2, R_2, \dots)$. The agent uses the experience to recursively update Q-function $\{Q_t\}_{t \ge 1}$, $Q_t \colon \ALPHABET Z \times \ALPHABET A \to \reals$ as follows:
\begin{multline}\label{eq:ASQL}
    Q_{t+1}(z,a) = Q_t(z,a) \\ + \alpha_t(z,a) 
    \Bigl[ R_t + \gamma \max_{a' \in \ALPHABET A} Q_t(Z_{t+1},a') - Q_t(z,a) 
    \Bigr]
\end{multline}
where $R_t = r(S_t,A_t)$ is the reward at time~$t$ and $\{\alpha_t\}_{t \ge 1}$ are the learning rates which are chosen such that $\alpha_t(z,a) = 0$ when $(z,a) \neq (Z_t, A_t)$.

As mentioned earlier, the agent state does not satisfy the Markov property. So the standard proof of Q-learning for MDPs~\cite{watkins1992QL,Jaakkola1994,neuroDP} is not directly applicable. Nonetheless, it is shown in~\cite{Sinha2024} that ASQL converges under the following assumptions:
\begin{enumerate}
    \item [\textbf{(A1)}] 
    The behavioral policy $\mu$ is such that the Markov chain\footnote{As argued in \autoref{sec:product}, under any policy $\mu \in \PiZSS$ the product process $\{(S_t,Z_t)\}_{t \ge 1}$ is a Markov chain. This directly implies that the process $\{(S_t, Y_t, Z_t, A_t)\}_{t \ge 1}$ is a Markov chain as well.} $\{(S_t, Y_t, Z_t, A_t)\}_{t \ge 1}$ is irreducible and aperiodic and, therefore, has a limiting distribution $\zeta^{\mu}$. Moreover, under the behavioral policy, all $(z,a)$ are visited infinitely often, i.e., $\sum_{(s,y)}\zeta^{\mu}(s,y,z,a) > 0$.
    \item [\textbf{(A2)}]
    The learning rates $\{\alpha_t\}_{t \ge 1}$ satisfy the standard Robbins–Monro conditions, i.e., for all $(z,a)$ we have that
    \[
        \sum_{t \ge 1} \alpha_t (z,a) = \infty
        \quad\text{and}\quad
        \sum_{t \ge 1} \alpha_t (z,a)^2 < \infty, \space a.s.
    \]
\end{enumerate}

For the ease of notation, we will continue to use $\zeta^{\mu}$ to denote the marginal and conditional distributions w.r.t.\ $\zeta^{\mu}$. In particular, for marginals we use $\zeta^{\mu}(y,z,a)$ to denote $\sum_{s \in \ALPHABET S} \zeta^{\mu}(s,y,z,a)$ and so on; for conditionals, we use  $\zeta^{\mu}(s | z,a)$ to denote $\zeta^{\mu}(s,z,a) / \zeta^{\mu}(z,a)$ and so on.

We now state the main result of~\cite{Sinha2024}.\footnote{The result proved in~\cite{Sinha2024} is a generalization of \autoref{thm:asql} to periodic agent-state based policies. Taking a period of $1$ implies \autoref{thm:asql}.}
\begin{colortheorem}{Convergence of ASQL}{asql}
    Under assumptions (A1) and (A2), the iterates $\{Q_t\}_{t \ge 1}$ of ASQL~\eqref{eq:ASQL} converge almost surely to a limit $Q^{\mu}_{\ASQL}$, which is the fixed point of the following DP:
    \begin{multline}
        Q^{\mu}_{\ASQL}(z,a) = r^{\mu}_{\ASQL}(z,a) 
        \\ + \gamma 
        \sum_{z' \in \ALPHABET Z} P^{\mu}_{\ASQL}(z'|z,a) \max_{a' \in \ALPHABET A} Q^{\mu}_{\ASQL}(z',a')
    \end{multline}
    where
    \begin{align*}
        r^{\mu}_{\ASQL}(z,a) &= \sum_{s \in \ALPHABET S} \zeta^{\mu}(s | z, a) r(s,a), \\
        P^{\mu}_{\ASQL}(z'|z,a) &= \smashoperator{\sum_{(s,y') \in \ALPHABET S \times \ALPHABET Y} }
        \IND_{\{ z' = \phi(z,y',a) \}} P(y'|s,a) \zeta^{\mu}(s|z,a).
    \end{align*}
\end{colortheorem}

Let $\pi^{\mu}_{\ASQL} \in \PiZSD$ denote the policy
\begin{equation}
    \pi^{\mu}_{\ASQL}(z) \coloneqq \arg\max_{a \in \ALPHABET A} Q^{\mu}_{\ASQL}(z,a)
\end{equation}
\autoref{thm:asql} implies that the greedy policy corresponding to $\{Q_t\}_{t \ge 1}$ computed via~\eqref{eq:ASQL} converges to $\pi^{\mu}_{\ASQL}$. Define $\boldsymbol{\vec \pi}^{\mu}_{\ASQL} = (\vec \pi^{\mu}_{\ASQL,1}, \vec \pi^{\mu}_{\ASQL,2}, \dots)$ to be the history dependent policy given by 
\begin{equation}
    \vec \pi^{\mu}_{\ASQL,t}(h_t) = \pi^{\mu}_{\ASQL}(\sigma_t(h_t)).
\end{equation}

A natural follow-up question is: How well does $\boldsymbol{\vec \pi}^{\mu}_{\ASQL}$ perform? The answer to this question is surprisingly easy because we can view $(\boldsymbol{\vec \sigma}, P^{\mu}_{\ASQL}, r^{\mu}_{\ASQL})$ as an AIS. Therefore, \autoref{thm:ais} provides an immediate bound on $\JHND - J^{\boldsymbol{\vec \pi}^{\mu}_{\ASQL}}$ as follows:
\begin{corollary}\label{cor:asql}
   We have that
   \[
     \JHND - J^{\boldsymbol{\vec \pi}^{\mu}_{\ASQL}} 
     \le
     \frac {2}{1-\gamma} 
     \bigl[ \varepsilon + \gamma \delta \rho_{\F}(V^{\mu}_{\ASQL}) \bigr]
   \]
   where $V^{\mu}_{\ASQL}(z) \coloneqq \max_{a \in \ALPHABET A} Q^{\mu}_{\ASQL}(z,a)$ and $\varepsilon$ and $\delta$ are defined in a manner similar to \autoref{sec:ais}
\end{corollary}

The results of \autoref{thm:asql} and \autoref{cor:asql} suggest that the performance of ASQL could be improved by choosing the state update function $\phi$ so as to minimize the approximation losses $\varepsilon$ and $\delta$. In particular, consider an implementation of Q-learning where a recurrent neural network (RNN) such as LSTM (long short term memory) or GRU (gated recurrent unit) is used as a history compression function $\boldsymbol{\vec \sigma}$. Add an ``AIS-block'' which learns a generative model for $(P_{\ASQL}, r_{\ASQL})$ by using  the \emph{AIS loss} $\lambda \varepsilon^2 + (1-\lambda) \delta^2$ as an auxiliary loss. This algorithm is called RQL-AIS in~\cite{EWRL-RQL}. Note that we can generate RQL-AIS version of any implementation of recurrent Q-learning for POMDPs.

In~\cite{EWRL-RQL}, a version of RQL-AIS using R2D2~\cite{kapturowski2018recurrent} as a base implementation is compared with R2D2 on the minigrid benchmark~\cite{minigrid,minigrid-paper}. The results of~\cite{EWRL-RQL} show that adding an AIS block significantly improve the performance of Q-learning in the larger and more complicated minigrid environments.

\subsection*{Remarks on ASQL}
\begin{enumerate}
    \item A key feature of \autoref{thm:asql} is that the limit $Q^{\mu}_{\ASQL}$ depends on the behavioral policy $\mu$, unlike in MDPs where the limit is policy-independent (as long as all state-action pairs are visited infinitely often). This dependence arises because the agent state is not an information state and thus is not policy-independent. See \cite{witsenhausen1975policy} for a discussion on policy independence of information states.
    
    \item  The idea of adding \emph{AIS losses} as an auxiliary loss for ASQL was first proposed in~\cite{subramanian2022approximate} for actor-critic algorithms. It is argued in~\cite{ni2024bridging} that in ASQL it is not useful to learn the reward function $r^{\mu}_{\ASQL}$ since the Q-function is already being learnt. Rather one should simply use $\delta^2$ as an auxiliary loss. 
    \item Learning a dynamics model to minimize $\delta^2$ is effectively the same as self-supervised representation learning, which has been used as a heuristic in different forms in MDPs and POMDPs in the literature. See~\cite{ni2024bridging} for a detailed discussion.
    
    \item The convergence of the simplest version of ASQL was established in~\cite{singh1994learning} for $Z_t = Y_t$ under the assumption that the actions are chosen i.i.d.\ (and do not depend on $Z_t$). In~\cite{Perkins2002} it was established that $Q^{\mu}_{\ASQL}$ is the fixed point of~\eqref{eq:ASQL}, but convergence of $\{Q_t\}_{t \ge 1}$ to $Q^{\mu}_{\ASQL}$ was not established. The convergence of ASQL when the agent state is a finite window of past observations and actions was established in~\cite{Kara2022}. The convergence result for a general agent state presented in \autoref{thm:asql} follows essentially the same proof argument.

    \item ASQL is closely related to Q-learning for MDPs with state aggregation or quantization~\cite{Singh1994aggregation,tsitsiklis1996feature,kara2021q}. In particular, consider an MDP with large or continuous state space, where we partition the state space into bins, and treat each bin as an aggregated state. This model is equivalent to a PODMP where the observation equals the index of the bin corresponding to the current state. Thus, Q-learning in MDPs with state aggregation is equivalent to ASQL with agent state equal to current observation. Therefore, there are considerable similarities between the convergence analysis and approximation bounds of these two setups.

    \item ASQL may also be viewed as a MDP with non-Markovian state~\cite{chandak2024reinforcement,Kara2024}. As such there are considerable similarities between the convergence analysis and approximation bounds for these two setups.
    
    \item The regret of an optimistic variant of ASQL is presented in~\cite{Dong2022}, though the exact setting there is slightly different. See~\cite{Dong2022} for details.
\end{enumerate}

\subsection{Agent-state based actor critic (ASAC)}

ASQL always learns a deterministic policy (i.e., a policy belonging to $\PiZSD$). As highlighted by~\eqref{eq:diagram} and fact (F2), stochastic policies can perform better than deterministic policies. One approach to learn stochastic policies is to use the actor-critic class of algorithms. 

An agent-state based actor critic (ASAC) algorithm was proposed in~\cite{subramanian2022approximate}. We present it with a slightly different perspective here. Consider any policy $\pi \in \PiZSS$. Motivated by the result of \autoref{thm:asql}, we can run a temporal difference learning algorithm to learn a Q-function corresponding to $\pi$ as follows:
\begin{multline}\label{eq:ASAC}
    Q^{\pi}_{t+1}(z,a) = Q^{\pi}_t(z,a) \\ + \alpha_t(z,a) 
    \Bigl[ R_t + \gamma Q^{\pi}_t(Z_{t+1},A_{t+1}) - Q^{\pi}_t(z,a) 
    \Bigr]
\end{multline}
where $R_t = r(S_t,A_t)$ as before. \autoref{thm:asql} implies that under conditions similar to (A1) and (A2), the iteration~\eqref{eq:ASAC} converges almost surely to a limit $Q^{\pi}_{\ASAC}$, which is the fixed point of the following DP:
\begin{multline}
    Q^{\pi}_{\ASAC}(z,a) = r^{\pi}_{\ASAC}(z,a) 
    \\ + \gamma 
    \smashoperator{\sum_{(z',a') \in \ALPHABET Z \times \ALPHABET A}} \pi(a'|z') P^{\pi}_{\ASAC}(z'|z,a) Q^{\pi}_{\ASAC}(z',a')
\end{multline}
where
\begin{align*}
    r^{\pi}_{\ASAC}(z,a) &= \sum_{s \in \ALPHABET S} \zeta^{\pi}(s,a) r(s,a), \\
    P^{\pi}_{\ASAC}(z'|z,a) &= \smashoperator{\sum_{(s,y') \in \ALPHABET S \times \ALPHABET Y} }
    \IND_{\{ z' = \phi(z,y',a) \}} P(y'|s,a) \zeta^{\pi}(s|z,a).
\end{align*}

Define $J^{\pi}_{\ASAC} = \sum_{(z,a) \in \ALPHABET Z \times \ALPHABET A} \xi_1(z,a) Q^{\pi}_{\ASAC}(z,a)$.
Moreover, define $\boldsymbol{\vec \pi} = (\vec \pi_1, \vec \pi_2, \dots)$ to be the history dependent policy given by 
\[
    \vec \pi_t(h_t) = \pi(\vec \sigma_t(h_t)).
\]
Then, by an argument similar to \autoref{thm:ais} we have that (see~\cite{subramanian2022approximate})
\[
    \bigl| J^{\boldsymbol{\vec \pi}} - J^{\pi}_{\ASAC} \bigr|
    \le 
    \frac{1}{1-\gamma} 
    \bigl[ \varepsilon + \gamma \delta \rho_{\F}(V^{\pi}_{\ASAC}) \bigr]
\]
where $V^{\pi}_{\ASAC}(z) = \max_{a \in \ALPHABET A} Q^{\pi}_{\ASAC}(z,a)$ and $\varepsilon$ and $\delta$ are defined in a manner similar to \autoref{sec:ais}. 

The above discussion suggests the following algorithm. Consider an implementation of an actor-critic algorithm where an RNN such as LSTM or GRU is used as a history compression function $\boldsymbol{\vec \sigma}$. Similar to RQL-AIS, add an ``AIS block'' which learns a generative model for $(P_{\ASAC}, r_{\ASAC})$ by using $\lambda \varepsilon^2 + (1-\lambda) \delta^2$ as an auxiliary loss. This algorithm is called PORL-AIS in~\cite{subramanian2022approximate}. Note that we can generate a PORL-AIS version of any implementation of recurrent actor-critic for POMDPs.

In~\cite{subramanian2022approximate}, a version of PORL-AIS is compared with LSTM-based actor critic on the minigrid benchmark~\cite{minigrid,minigrid-paper}. The results of~\cite{subramanian2022approximate} show that adding an AIS block significantly improves the performance of actor-critic algorithms in the larger and more complicated minigrid environments.

\subsection*{Remarks of ASAC}
\begin{enumerate}
    \item It is also possible to add AIS losses to actor only algorithms such as G(PO)MDP~\cite{Baxter2001}. See~\cite{subramanian2022approximate} for details.

    \item We have argued that the TD iterations for $\{Q_t\}_{t \ge 1}$ converge to a limit. Therefore, for a parameterized policy $\pi_{\theta}$, the gradient computed via the policy gradient formula
    \[
        \frac 1T \sum_{t=1}^T \GRAD_{\theta} \log \pi_{\theta}(A_t | Z_t) Q^{\pi_{\theta}}_t(Z_t, A_t)
    \]
    asymptotically converges to 
    \[
        \sum_{(z,a) \in \ALPHABET Z \times \ALPHABET A} \zeta^{\pi}(z,a) \log \pi_{\theta}(a|z) Q^{\pi}_{\ASQL}(z,a)
    \]
    which is an unbiased estimate of $\GRAD_{\theta} J^{\pi_{\theta}}_{\ASAC}$ rather than $\GRAD_{\theta} J^{\boldsymbol{\vec \pi}_{\theta}}$. Therefore, PORL-AIS does not have the local convergence guarantees of actor critic for MDPs (but neither do other actor-critic algorithms for POMDPs!).

    \item AIS error bounds for asymmetric actor critic~\cite{Baisero2022} are developed in~\cite{Sinha2023} but the numerical experiments presented there suggest that adding an AIS block to asymmetric actor critic algorithms does not provide any measurable performance improvement over the vanilla asymmetric actor critic algorithm.
\end{enumerate}

\section{Conclusion}\label{sec:conclusion}

In this tutorial, we have summarized some of the approaches used for planning and learning agent-state based policies in POMDPs. Our survey is not exhaustive. There are other alternative approaches which we have not discussed (e.g., predictive state representations~\cite{Boots2011,Hamilton2014,Jiang2016,Kulesza2015}, bisimulation~\cite{CastroPanangadenPrecup_2009,castro2021mico}, world models~\cite{werbos1974beyond,schmidhuber1990making,worldmodels,Hafner2020Dream,Hafner2021mastering}, decision transformers~\cite{decisiontransformer,ni2023when}). Nonetheless, it is our hope that this paper will present the reader with a unified perspective on a subset of the literature. We conclude by emphasizing that this is a rapidly developing research area and many of the fundamental questions are still open. 

\section*{Acknowledgment}

The authors are grateful to Demosthenis Teneketzis (University of Michigan), Jayakumar Subramanian (Adobe Inc.), and Matthieu Geist (Cohere) for discussions and collaborations which helped the authors develop the viewpoint presented in this tutorial. They also thank Charalambos Charalambous (University of Cyprus), Vijay Subramanian (University of Michigan), and Serdar Y\"uksel (Queen's University), and Tianwei Ni (University of Montreal) for helpful feedback..

\bibliographystyle{IEEEtranS}
\bibliography{IEEEabrv,references}

\end{document}